\documentclass[fleqn,twoside]{article}
\usepackage{espcrc2}

\usepackage{graphicx}
\usepackage[figuresright]{rotating}

\newcommand{\AmS}{{\protect\the\textfont2
  A\kern-.1667em\lower.5ex\hbox{M}\kern-.125emS}}

\title{Phase transition strengths from the density of partition 
       function zeroes}

\author{Wolfhard Janke\address[ITP]{Institut f\"ur Theoretische Physik, 
        Universit\"at Leipzig,
        Augustusplatz 10/11, 04109 Leipzig, Germany}
        and
        Ralph Kenna\address[TCD]{School of Mathematics, 
        Trinity College Dublin, Ireland}
        }
 
\begin{document}

\begin{abstract}
We report on a new method to extract thermodynamic properties from the 
density of partition function zeroes on finite lattices. This allows direct 
determination of the order and strength of phase transitions numerically. 
Furthermore, it enables efficient distinguishing between first- and 
second-order transitions, elucidates crossover between them and illuminates 
the origins of finite-size scaling. The power of the method is illustrated in
typical applications for both Fisher and Lee-Yang zeroes.
\end{abstract}

\maketitle

\section{INTRODUCTION}
\label{Introduction}

The characterisation of  phase transitions, in particular of their
order and strength, is among the hard numerical problems that are
common to lattice field theory and spin model physics. Frequently
applied techniques focus either on the finite-size scaling (FSS) behaviour of
thermodynamic functions such as the specific heat, susceptibility or
Binder parameter, or somewhat more ``microscopically'' on the limiting shape of
the underlying probability densities of energy and magnetization as the
thermodynamic limit is approached. A related and increasingly popular 
alternative approach are FSS analyses of zeroes of the partition function
\cite{IPZ}.

If $t=T/T_c-1$ denotes the reduced temperature and $h$ the external field,
then the FSS of the $j^{\rm{th}}$ complex partition function zero 
for a $d$-dimensional system of linear extent $L$ is given by
\begin{equation}
 t_j(L) \sim \left( {j}/{L^d} \right)^{1/\nu d}
\quad, 
\label{FSSa}
\end{equation}
\begin{equation}
 h_j(L) \sim \left({j}/{L^d}\right)^{(d+2-\eta)/2d}
\quad,
\label{FSSb}
\end{equation}
where $\nu$ and $\eta$ are the standard critical exponents.
In (\ref{FSSa}) we assume $h=0$ and
$t_j(L)$ are called Fisher zeroes. Conversely, in (\ref{FSSb}),
$t=0$ is assumed and $h_j(L)$ are the Lee-Yang zeroes.
The standard approach to FSS of zeroes is to fix the index to
$j=1$ and extract an estimate for the critical
exponents from a range of lattice sizes.

In recent years, however, there have also been some attempts \cite{density} 
to extract
the density of zeroes (a continuous function) from their (discrete)
distribution for a finite and numerically accessible lattice.
In view of the increasing importance attached to this approach, we
recently suggested an appropriate way this should be done \cite{us}. 

\section{DENSITY OF ZEROES}
\label{theory}

The partition function for finite $L$ is
$ Z_L(z) \propto  \prod_{j}{\left(z-z_j(L)\right)}$,
where $z$ is an appropriate function of temperature
or field. We assume the zeroes, $z_j$, are 
on a line impacting on to the real axis 
at the critical point, $z_c$. 
Parameterising zeroes on this  line by
$z_j=z_c+r_j \exp{(i \varphi)}$ we may define 
the density of zeroes as
$
 g_L(r) = L^{-d} \sum_{j} \delta(r - r_j(L))
$.
The cumulative distribution function of zeroes is then
$
 G_L(r)
 =
 \int_0^r{ g_L(s) d s}
$
which is $j/L^d$ if $ r \in (r_j,r_{j+1})$.
At a zero one may assume the cumulative density is given by the average
$
 G_L(r_j) =  (2j-1)/2L^d
$.

For a first-order phase transition this integrated density of zeroes is, in 
the thermodynamic limit, given by
\begin{equation}
 G_\infty(r) = g_\infty(0) r 
\quad ,
\label{1st}
\end{equation}  
so that the density is non-vanishing at the real axis  \cite{LY}.
The slope at the origin in (\ref{1st})  is related to the latent heat
(magnetization) in the Fisher (Lee-Yang) case via \cite{LY}
$ g_\infty(0) \propto  \Delta e$. 

For a second-order transition
the corresponding expressions for Fisher and Lee-Yang zeroes are \cite{Abe}
\begin{equation}
 G_\infty(r) \propto  r^{2-\alpha} 
\quad\! {\rm{and}}\! \quad 
G_\infty(r) \propto  r^{2d/(d+2-\eta)}
\quad \!\!\!.
\label{2nd}
\end{equation}

Traditional FSS emerges quite naturally from this density approach.
Equating $G_L(r_j)$ to (\ref{2nd}) in the second-order Fisher case, gives 
the usual FSS formula for fixed index zeroes, $r_j(L) \sim L^{-1/\nu}$,
where $r_j$ may be taken to be the imaginary part of the $j^{\rm{th}}$ zero. 
Similarly, in the Lee-Yang case, one recovers the fixed index FSS formula 
$h_j(L) \sim L^{-(d+2-\eta)/2}$. Moreover, considering (\ref{1st}) gives 
$r_j(L) \sim L^{-d}$, explaining also the usual identification of $\nu$ with
$1/d$ for a first-order temperature driven phase transition.

A plot of $G_L(r_j)$ against $r_j(L)$
should thus ({\em{i}\/}) go through the origin, ({\em{ii}\/}) display $L$- and
$j$- collapse and ({\em{iii}\/}) reveal the order and strength of the 
phase transition by its slope near the origin.

\section{APPLICATIONS}
\label{applications}

Superimposing the behaviour (\ref{1st}) and (\ref{2nd}) at 
first- and second-order transitions, the ansatz for  
the cumulative density can be written as 
\begin{equation}
 G(r) = a_1 r^{a_2} + a_3
\quad ,
\label{gen}  
\end{equation}
where we also introduced an additional parameter $a_3$ signalising
the absence of a phase transition: if $a_3 > 0$ the zeroes have already 
crossed the real axis (broken phase scenario) while
for $a_3<0$ the zeroes have not yet reached the real axis 
(symmetric phase). For Fisher zeroes, a first-order transition is 
indicated if $a_2 \sim 1$ for small $r$, in which case
the latent heat is proportional to the slope $a_1$.
A value of $a_2$ larger than $1$ signals a second-order 
transition whose strength is given by $\alpha = 2-a_2$.


%
\begin{figure}[t]
\vspace{10cm}
\includegraphics{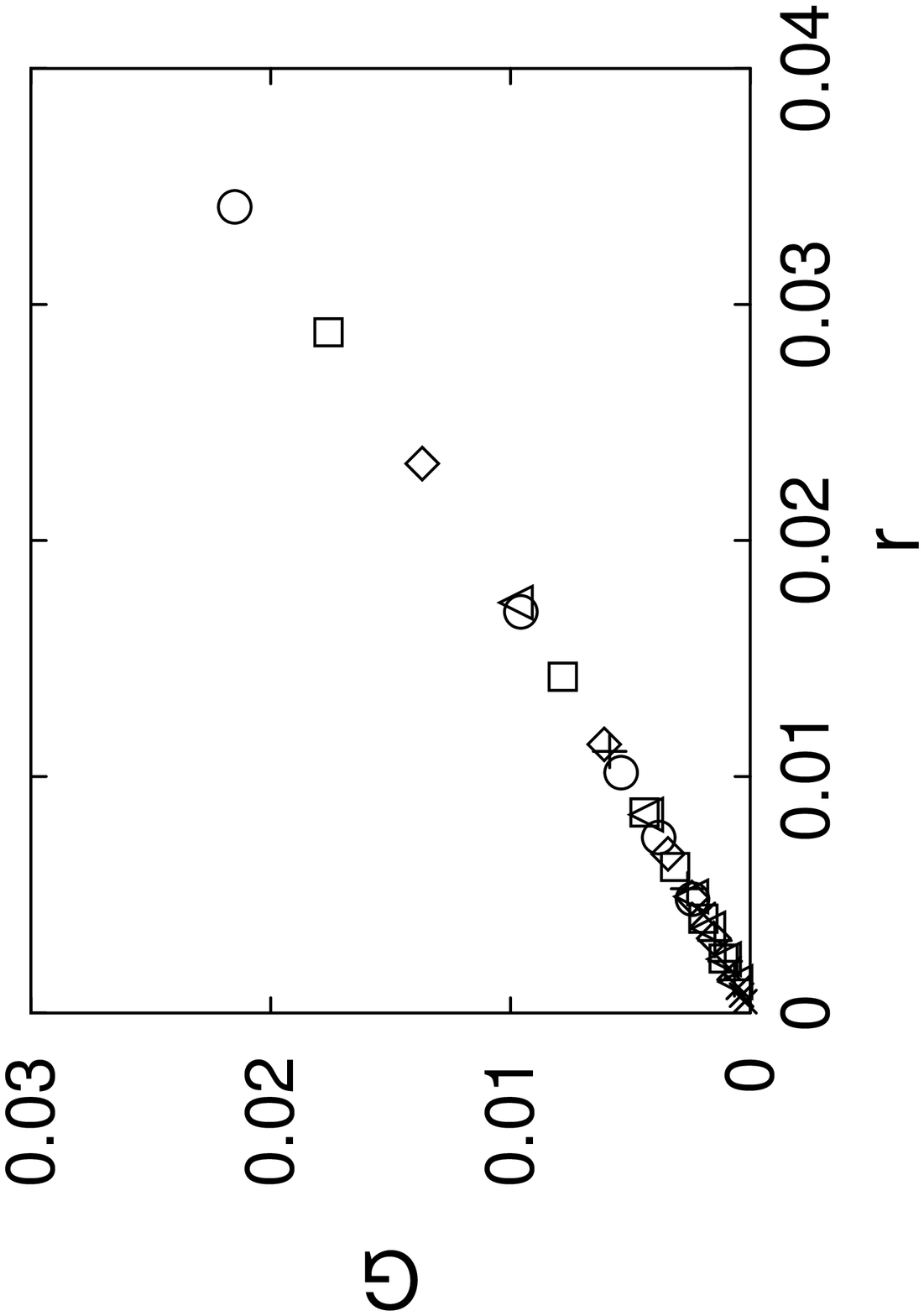}
\includegraphics{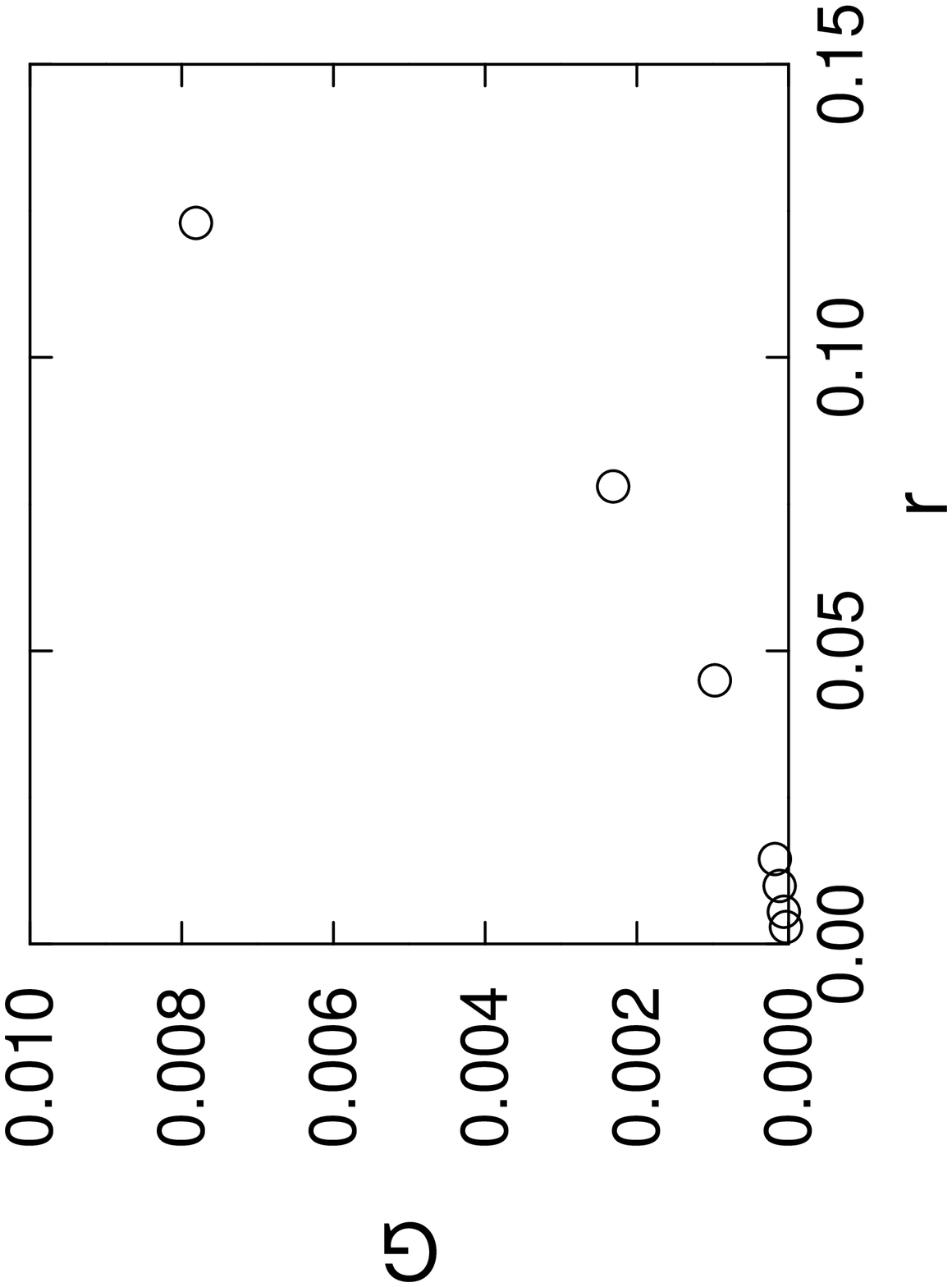}
\includegraphics{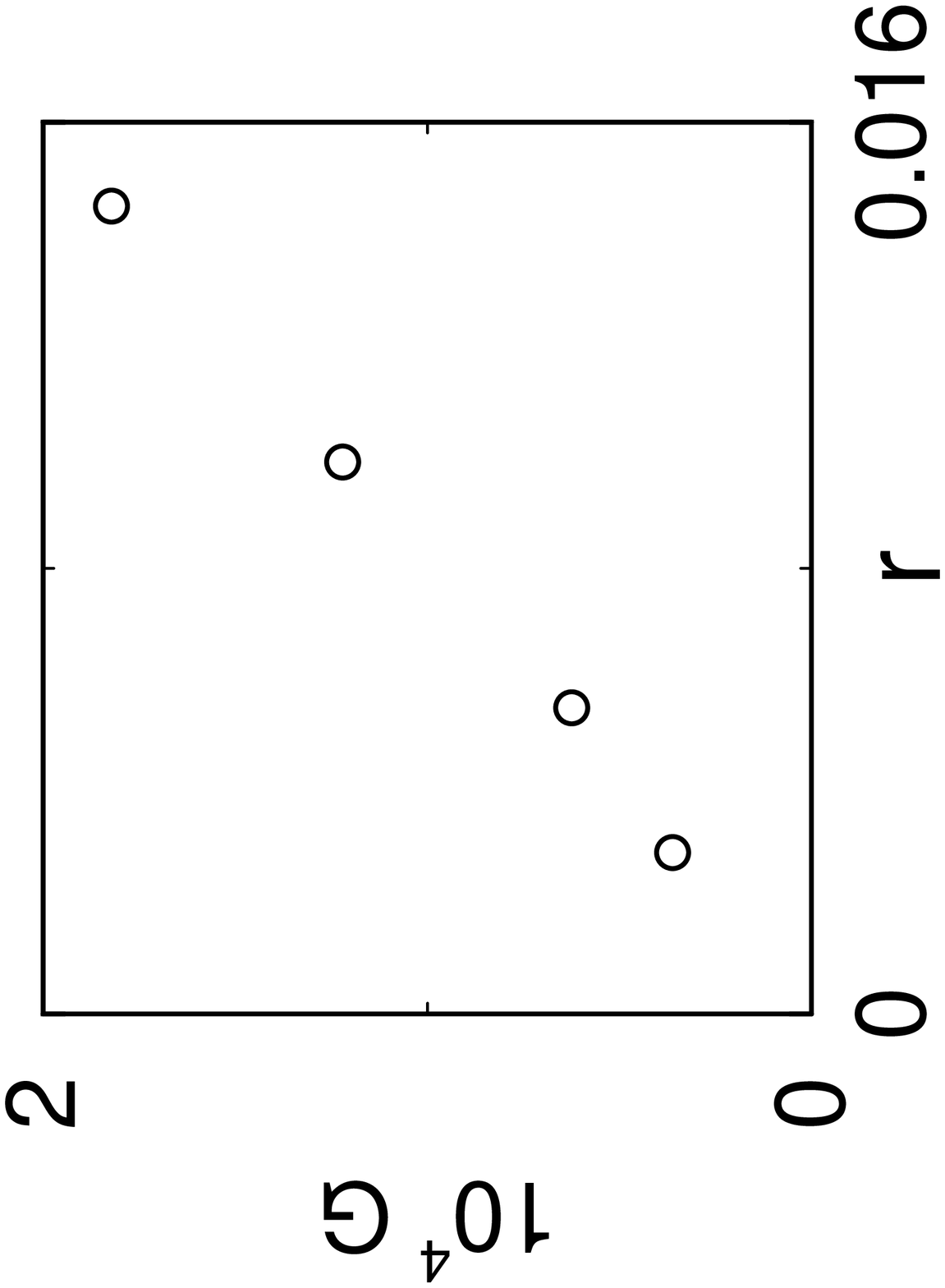}
\caption[a]{Distribution of partition function zeroes.
(a) 2D 10-state Potts model and (b) 3D $L_t=4$ SU(3) lattice gauge theory.
\begin{picture}(0,0)
\put(32,230){\bf (a)}
\put(32,74){\bf (b)}
\end{picture}
\vspace*{-0.3cm}
}
\label{fig:d=2.q=10}
\end{figure}

\begin{figure}[t]
\vspace{4.5cm}
\includegraphics{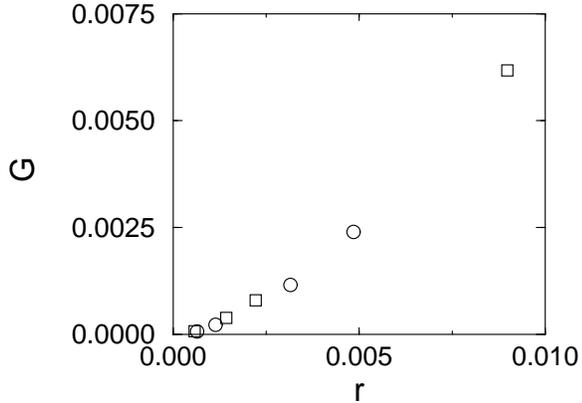}
\caption[a]{4D Abelian surface gauge model. 
}
\label{fig:d=4.surface.gauge}
\end{figure}

\noindent {\bf 2D 10-State Potts Model:} 
This is the paradigm for models exhibiting a strong first-order transition.
Using the first six Fisher zeroes for $L=4$--$64$ as listed
in \cite{Vi91} we find the distribution of zeroes depicted in
Fig.~\ref{fig:d=2.q=10}(a). The
excellent data collapse for various $L$ and $j$
indicates that the interpolated $G_L(r_j)$ is the proper choice.
Fitting (\ref{gen}) to the $L=16$--$64$, $j=1$--$4$
data points gives $a_2=1.10(1)$ and $a_3= 0.00004(1)$, a strong
indication of a first-order transition.  Fixing $a_3=0$, $a_2=1$,
a single-parameter fit close to the origin yields a slope corresponding 
to latent heat $\Delta e = 0.698(2)$
which compares well with the exact value of $0.6961$.


\noindent {\bf 3D SU(3) Lattice Gauge Theory:} 
Here we consider the deconfinement transition
for $L_t L^3$ lattices. The lowest Fisher zeroes for $L_t=4$
and spatial extent $L=4$--$24$ are given in \cite{AlBe92}.
Applying standard FSS analysis to the $L \ge 14$ data only
yields $\nu = 0.35(2)$, compatible with $1/d = 0.33$ and
thus indicative of a first-order transition, while fits for
$L \le 8$ suggest a continuous transition.
Figure~\ref{fig:d=2.q=10}(b) shows the distribution of zeroes for all 
lattices, and the insert highlights $L \ge 14$.
The figure, clearly supportive of a non-zero slope through the origin,
justifies restricting the analysis to the largest lattices
and thereby elucidating the
procedure of deciding where FSS sets in.
This slope is $0.0121(3)$,  implying a latent heat of $0.0760(19)$
in agreement with the estimate $0.0758(14)$ using
standard methods \cite{AlBe92}.


\noindent {\bf 4D  Abelian Surface Gauge Model:} 
Being the dual of the 4D Ising model one expects for this model, up to 
logarithmic corrections,
mean-field critical exponents $\alpha=0$, $\nu = 1/2$. The first two Fisher 
zeroes for lattices of size $L=3$--$12$ are listed in \cite{BaVi94} where
a conventional analysis applied to the first index zero
yields the best estimate of $\nu = 0.469(17)$ 
from the two largest lattices. 
A fit of (\ref{gen}) to the distribution in Fig.~\ref{fig:d=4.surface.gauge} 
yields $a_2$ incompatible with unity. Using the data near the origin gives 
$a_2=1.90(9)$ or $\alpha = 0.10(9)$, compatible with zero. 

\begin{figure}[t]
\vspace{4.5cm}
\includegraphics{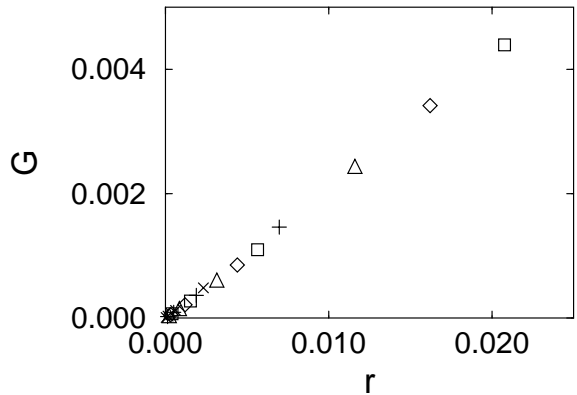}
\caption[a]{2D XY model (Lee-Yang case).
}
\label{fig:d=2.xy}
\end{figure}

\noindent {\bf 2D XY Model:} 
Here we demonstrate that the density technique is also applicable
in the Lee-Yang case. Figure~\ref{fig:d=2.xy} depicts the distribution of
these zeroes for the 2D XY model at the
critical point, $\beta_c = 1.113$, obtained for lattice sizes $L=32$--$256$
\cite{KeIr97}. From (\ref{2nd}), and with $\eta = 1/4$,
one expects $G(r) \sim r^{16/15}$. A three-parameter fit (\ref{gen}) 
gives $a_3 = 0$, indicating that criticality has indeed been reached. A 
two-parameter fit now yields $a_2=1.063(3)$, compatible with expectation 
(taking logarithmic corrections into account).

%

\section{CONCLUSIONS}
\label{conclusions}  

We have discussed a new method to extract the (continuous) density of zeroes
from (discrete) finite-size data and demonstrated how this can be used to 
distinguish between phase transitions of first and second order as well as
to measure their strengths. The method meets with a high degree of success in 
lattice field theory and statistical physics and lends new insights into the
origins of finite-size scaling.


\end{document}